\pdfoutput=1

\documentclass[iop,apjl]{emulateapj}

\newdimen\hssize
\usepackage{amsmath}
\usepackage{framed}
\usepackage{txfonts}
\usepackage{epstopdf}
\usepackage{color}
\usepackage{rotating}
\usepackage{natbib}
\usepackage{ulem}
\usepackage{xspace}

\newcommand{\mpch}{\>h^{-1}{\rm {Mpc}}}

\newcommand{\msun}{\>{\rm M_{\odot}}}

\def\gcm3{\mathrm{g} / \mathrm{cm}^3}

\def\gtsima{$\; \buildrel > \over \sim \;$}
\def\ltsima{$\; \buildrel < \over \sim \;$}
\def\prosima{$\; \buildrel \propto \over \sim \;$}
\def\gsim{\lower.7ex\hbox{\gtsima}}
\def\lsim{\lower.7ex\hbox{\ltsima}}
\def\simgt{\lower.7ex\hbox{\gtsima}}
\def\simlt{\lower.7ex\hbox{\ltsima}}
\def\simpr{\lower.7ex\hbox{\prosima}}

\newcommand{\avg}[1]{\langle #1 \rangle}

\def\rmd{{\rm d}}

\def\rml{{\rm l}}
\def\rmm{{\rm m}}

\def\rmp{{\rm p}}

\def\rms{{\rm s}}

\def\rmA{{\rm A}}

\slugcomment{To be submitted to the Astrophysical Journal}
\shorttitle{Cosmological dependence of observables from redshift surveys}
\shortauthors{More, S.}

\begin{document}


\def\figdir{.}
\def\figext{pdf}


\title{Cosmological dependence of the measurements of luminosity function, projected clustering and galaxy-galaxy lensing signal}
\author{Surhud More \altaffilmark{1,2}}
\affil{
$^1$ Kavli Institute for the Physics and Mathematics of the Universe (WPI), The University of Tokyo, 5-1-5 Kashiwanoha, Kashiwa-shi, Chiba, 277-8583, Japan \\
$^2$ surhud.more@member.ipmu.jp
}

\begin{abstract}
Observables such as the luminosity function of galaxies, $\Phi(M)$, the
projected clustering of galaxies, $w_\rmp(r_\rmp)$, and the galaxy-galaxy
lensing signal, $\Delta\Sigma$, are often measured from galaxy redshift surveys
assuming a fiducial cosmological model for calculating distances to and between
galaxies. There is a growing number of studies that perform joint analyses of
these measurements and constrain cosmological parameters. We quantify the
amount by which such measurements systematically vary as the fiducial cosmology
used for the measurements is changed, and show that these effects can be
significant at high redshifts ($z\sim0.5$). We present a simple way that maps
the measurements made using a particular fiducial cosmological model to any
other cosmological model. Cosmological constraints (or halo occupation
distribution constraints) that use the luminosity function, clustering
measurements and galaxy-galaxy lensing signal but ignore these systematic
effects may underestimate the confidence intervals on the inferred parameters.
\end{abstract}

\keywords{cosmology: observations --- dark matter --- large-scale structure of
universe --- galaxies: distances and redshifts}


\section{Introduction}
\label{sec:intro}

Large scale galaxy redshift surveys such as the Sloan digital Sky Survey (SDSS
hereafter) have revolutionized the field of galaxy formation and cosmology.
Data from such surveys has enabled precise measurements of the abundance of
galaxies \citep[see e.g.,][]{Blanton2003}, the clustering of galaxies as a
function of luminosity \citep[see e.g.,][]{Zehavi2011} and the galaxy-galaxy
lensing signal \citep[see e.g.,][]{ Sheldon2004, Mandelbaum2006,
Mandelbaum2013}. These measurements have been used to constrain an important
outcome of the galaxy formation processes, the relation between galaxy
luminosity (or stellar mass) and the underlying halo mass
\citep{Cacciato2009,Cacciato2013,Leauthaud2012,Tinker2013}. It has also been
argued that a joint analysis of these measurements can be used to constrain
cosmological parameters, such as the matter density parameter and the amplitude
of cosmological fluctuations using the data on small scales \citep[see
e.g.,][]{Seljak2005,Yoo2006, Cacciato2009,vdb2013,More2013} and from large
scales \citep{Baldauf2010, Mandelbaum2013}. Cosmological constraints have been
obtained using the clustering of galaxies combined with other observables such
as the mass-to-light ratio on cluster scales \citep{vdb2003, Tinker2005,
vdb2007}, mass-to-number ratio \citep{Tinker2012} or the galaxy-galaxy lensing
signal \citep{Cacciato2013, Mandelbaum2013}.

To measure observables such as the luminosity function of galaxies, their
projected clustering signal and the galaxy-galaxy lensing, the galaxy angular
positions and redshifts are needed to be converted to cosmological distances
between us and the galaxies and between galaxies themselves. These conversions
are dependent upon the assumed cosmological model. It would be incorrect to
assume that the measurements do not change when the cosmological model used to
analyze the data is changed. Fitting analytical models to the same measurements
with varying cosmological parameters can affect the constraints derived on the
cosmological parameters. 

The objective of this short letter is to quantify this effect for the
particular set of observables, $\Phi(M)$, $w_\rmp(r_\rmp)$ and $\Delta\Sigma$,
and show that it is straightforward to account for such biases. In
Section~\ref{sec:analytics}, we quantify analytically the sensitivity of each
of these measurements to the assumed cosmological model. In
Section~\ref{sec:data}, we use the example of the projected galaxy clustering
signal to demonstrate that the effect on measurements of the changing reference
model can be accounted for in a simple manner. In Section~\ref{sec:summary}, we
summarize and discuss our results.


\section{Analytical estimates}
\label{sec:analytics}

\subsection{Galaxy luminosity function}
\label{sec:lf}

The abundance of galaxies is quantified by measurements of the luminosity
function of galaxies, $\Phi(M)\rmd M$, which gives the average number density
of galaxies within the absolute magnitude range $M\pm\rmd M/2$. The luminosity
functions are often determined from flux limited surveys and require us to
convert the apparent magnitude $m$, of a galaxy into an absolute magnitude, and
obtain the maximum distance out to which a galaxy with a given luminosity could
have been observed given the flux limit of the survey. These conversions are
dependent on the assumed cosmology in the following manner.

The apparent magnitude of a galaxy at redshift $z$ is related to the absolute
magnitude via the distance modulus $\mu(z)=5.0\log(D_{\rm
lum}[z,\Omega])+25$,
\begin{equation}
M=m-\mu(z)\,,
\end{equation}
where $D_{\rm lum}(z,\Omega)$ is the luminosity distance (in units of $\mpch$)
in a particular cosmological model, $\Omega$. The maximum comoving distance,
$\chi_{\rm max}$ to which this object can be observed given the magnitude limit
$m_{\rm lim}$ of the survey is given by
\begin{equation}
\chi_{\rm max}=\frac{1}{1+z}10^{0.2(m_{\rm lim}-M-25)}
\end{equation}
Thus differences in cosmology changes the luminosity of galaxies and the change
in $\chi_{\rm max}$ affects the normalization of the luminosity function. 

\begin{figure*}
\centering
\includegraphics{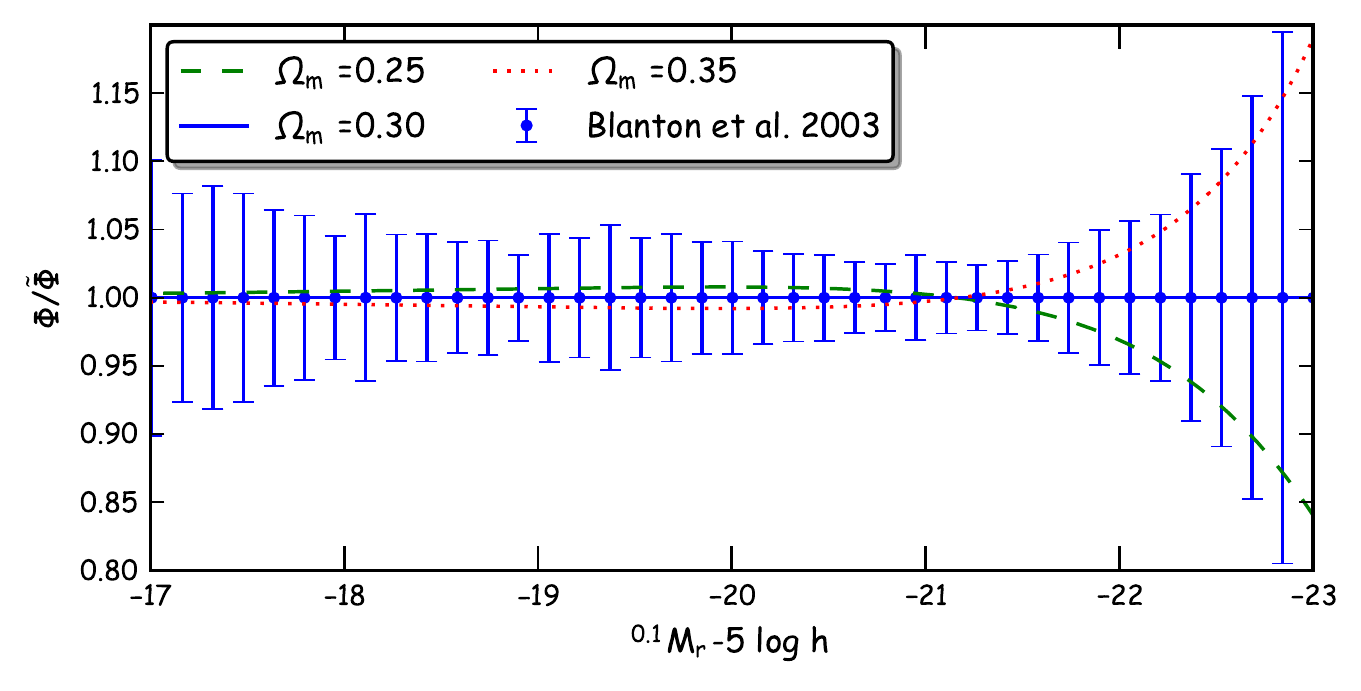}
\caption{
The ratio of the SDSS luminosity function from \citet{Blanton2003} derived in a
$\Omega_\rmm=0.3$ model with respect to the luminosity function expected if the
data was analyzed using $\Omega_\rmm=0.25$ or $\Omega_\rmm=0.35$ model.
}
\label{compare_lf}
\end{figure*}

To estimate the amount by which the luminosity function can change due to a
change in cosmological model, let us first assume that the luminosity function
as a function of redshift changes extremely weakly within the survey area used
to estimate it using some fiducial cosmological model \footnote{Of course, this
cannot be strictly true and is an assumption. If violated it is better to
divide up the survey in to finer redshift slices.}. The luminosity function
$\tilde\Phi(M,\tilde\Omega)$, measured in some fiducial cosmological model
$\tilde\Omega$ can be used to calculate the redshift dependence of the apparent
magnitude counts $N(m,z)\rmd m\rmd z$ per unit steradian, which is the
observable in a true sense,
\begin{equation}
N(m,z) \rmd m \rmd z = \tilde\Phi(m-\mu[z,\tilde\Omega]) \rmd m \tilde\chi^2
\frac{\rmd \tilde\chi}{\rmd z} \rmd z \,.
\end{equation}
These number counts can be reinterpreted as a luminosity function in a
cosmological model, $\Omega$ other than the fiducial model using the following
equation,
\begin{eqnarray}
\Phi(M,\Omega) \rmd M &=& \frac{1}{V} \int_{0}^{z_{\rm max}} N(M+\mu[z,\Omega],\,z)\,\rmd M\,\rmd z \\
&=& \frac{1}{V} \int_{0}^{z_{\rm max}} \tilde\Phi(M+\mu[z,\Omega]-\mu[z,\tilde\Omega])\,\rmd M\, \tilde\chi^2 \frac{\rmd \tilde\chi}{\rmd z} \rmd z. \nonumber \\
\end{eqnarray}
Here, $z_{\rm max}$ denotes the maximum redshift to which a galaxy can be
observed in the redshift survey, given its absolute magnitude in the
cosmological model $\Omega$, and $V$ denotes the comoving volume encompassed by
the Universe below this redshift. In Figure~\ref{compare_lf}, we use the
Schechter fit for the luminosity function provided by Blanton et al. 2003, in a
$\Omega_\rmm=0.3$ model and show the residuals of the luminosity function in
$\Omega_\rmm=0.25$ and $\Omega_\rmm=0.35$ models computed using the above
equation. We have assumed the magnitude limit in $r$-band of $17.77$
corresponding to the spectroscopic sample in SDSS.

Although the measurement errors on the luminosity function are large, the
difference is a systematic change in the shape of the luminosity function, and
can be as large as $\sim 20$ percent at the bright end. Since the errorbars are
difficult to propagate in an integral equation such as the one above, the ideal
way is to change the prediction for a particular cosmological model to predict
the counts in fiducial model used to obtain the luminosity function.

\begin{figure*}
\centering
\includegraphics{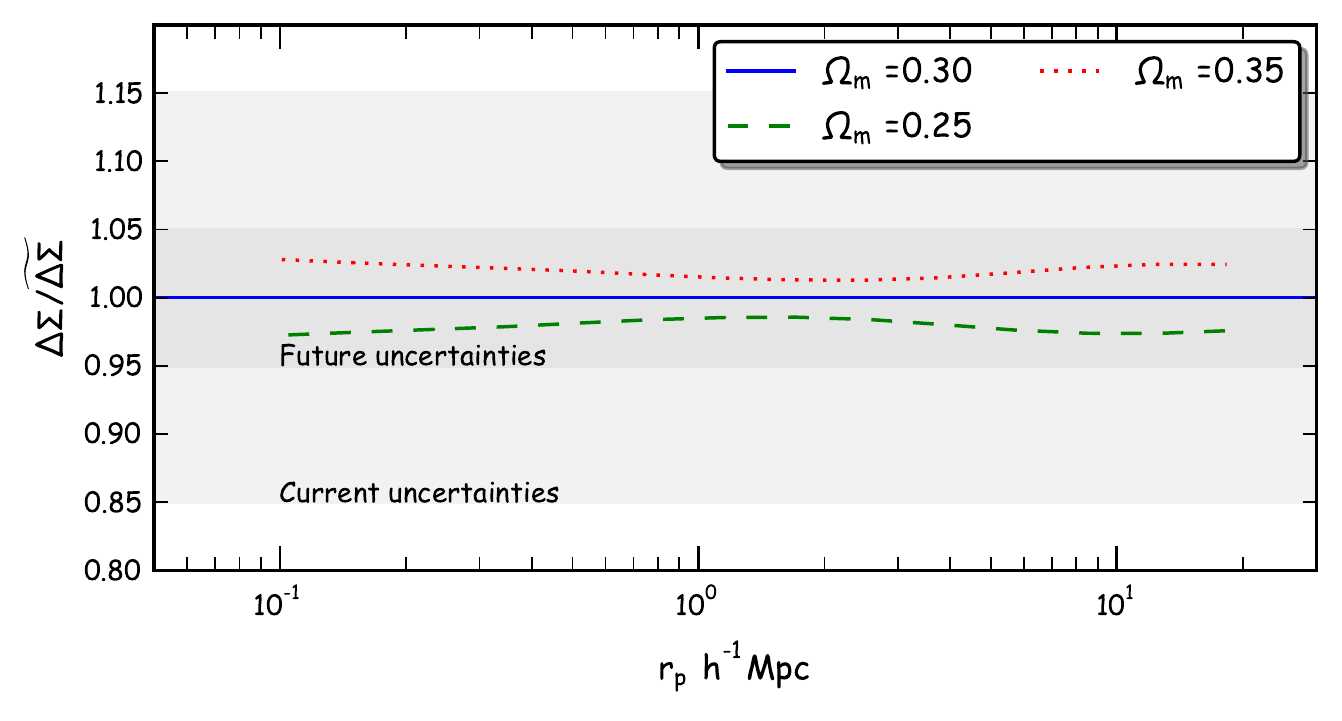}
\caption{
The ratio of the galaxy-galaxy lensing signal around BOSS galaxies expected
assuming an $\Omega_\rmm=0.3$ model with respect to that expected if the data
was analyzed using $\Omega_\rmm=0.25$ or $\Omega_\rmm=0.35$ model. The light
and dark grey band corresponds to current and future uncertainties.
}
\label{compare_esd}
\end{figure*}

\subsection{Projected clustering measurement}
\label{sec:wp}

For a flat $\Lambda$CDM model, the projected and the line-of-sight comoving
separations between two galaxies separated in redshift by a small difference
$\Delta z$ is given by,
\begin{equation}
r_\rmp = \chi(z_{\rm eff}) \theta; \,\,\,\, \pi=\frac{c}{H_0\,E(z_{\rm eff})} \Delta z\,\,,
\end{equation}
respectively, where $\chi(z_{\rm eff})$ is the comoving distance to the
effective redshift $z_{\rm eff}$, $\theta$ denotes the angular separation of
the two galaxies, $c$ denotes the speed of light, $H_0$
is the Hubble constant and $E(z)$ is the expansion function. Here we have
ignored the finite redshift width of the galaxy sample and assumed a particular
effective redshift to convert the angular positions and redshift into
distances.  These equations can then be used to count the number of pairs of
galaxies at a given separation vector $(r_\rmp,\pi)$, and compare it to the
number of pairs expected if the galaxies were distributed in a random manner.
The excess number of pairs over those expected in a random distribution gives
the clustering signal at the effective redshift, $\xi(r_\rmp,\pi)$. The
projected clustering is then obtained by integrating $\xi(r_\rmp,\pi)$ along
the line-of-sight,
\begin{eqnarray}
w_\rmp(r_\rmp,z_{\rm eff})&=&2\int_{0}^{\pi_{\rm max}} \xi(r_\rmp,\pi,z_{\rm eff}) \rmd\pi \,, \\
&=& 2\int_{0}^{\Delta z_{\rm max}} \xi(r_\rmp,\pi,z_{\rm eff})\frac{c}{H_0E(z_{\rm eff})} \rmd(\Delta z) \,.
\vspace{0.1cm}
\end{eqnarray}
In addition to changing slightly the composition of galaxy samples, there are
three different ways a change in the cosmology will affect the $w_\rmp$
measurements. If a cosmology other than the fiducial is used to analyze the
data, then  a given angular scale corresponds to a different comoving projected
separation scale. This difference can be small at low redshift \citep[see
e.g.][$\sim 1$ percent due to a change in $\Omega_\rmm$ from 0.25 to 0.3 at
$z=0.15$]{Zehavi2011}. As $w_\rmp \propto r_\rmp^{-1}$, the change in projected
separation scale roughly corresponds to a similar change in the value of
$w_\rmp(r_\rmp)$ when the cosmology is changed.

The second effect is due to change in the factor $E(z)$ as cosmology is
changed, and this changes the value of $w_\rmp(r_\rmp)$ by a multiplicative
constant at all scales. This effect is important, especially at higher redshift
(at $z\sim0.5$, the difference in the expansion functions is $3.5$ per cent
between $\Omega_\rmm$ of 0.25 and 0.3). Such effects due to the change of
transverse (and line-of-sight) scales are usually taken into account when
analysing baryon acoustic oscillations \citep[see
e.g.,][]{Blake2003,Eisenstein2005, Percival2007, Anderson2012} and redshift
space distortions \citep[see e.g.,][]{Bailinger1996, Tegmark2006, Blake2011}.

The third effect on $w_\rmp(r_\rmp)$ is subtle and is related to the change in
the integration limit in the above equation, as a given value of $\pi_{\rm
max}$ corresponds to a different value of $\Delta z_{\rm max}$. For
sufficiently large values of $\pi_{\rm max}$ as are employed in observations
(typically $60\sim100\mpch$) this effect is quite small as the value of
$\xi(r_\rmp,\pi)$ at large values of $\pi$ does not dominate the
$w_\rmp(r_\rmp)$ integral. Note however, that even this small difference can be
easily accounted for by adopting a different $\pi_{\rm max}$ value when
computing the analytical prediction.

In Table~\ref{table:compare}, we present the ratios of the comoving distance
and the expansion function for three different cosmological models, at
different redshifts. The range of cosmological models is chosen to be such that
it is well within the ranges of cosmological constraints obtained by a number
of joint analyses involving the clustering measurement. The fractions
$f^{\Omega_\rmm=0.30}_{E(z)}$ and $f^{\Omega_\rmm=0.30}_{\chi(z)}$ are defined
in the caption, and are chosen such that they roughly correspond to the
deviations in the clustering signal expected when the data is analysed in two
different cosmologies. The two effects change the clustering signal in the same
direction.

It can be seen that the combination of the first two effects are small $\sim 2$
per cent differences between $\Omega_\rmm=0.25$ and $\Omega_\rmm=0.35$ models
(although are systematically in the same direction on all scales) for low
redshift ($z\sim0.1$) analyses. At $z\sim0.5$, the effects cause $\sim 10$
percent differences between the two cosmological models above and can be very
important given the statistical errors in the measurements of $w_\rmp(r_\rmp)$
with current and upcoming large surveys.

\renewcommand{\tabcolsep}{0.5cm}
\begin{table}
\caption{Cosmological dependence of comoving distance and the expansion function}
\begin{center}
\begin{tabular}{cccc}
\hline\hline
Redshift & $\Omega_\rmm$ & $f^{\Omega_\rmm=0.3}_{\chi(z)}$ & $f^{\Omega_\rmm=0.3}_{E(z)}$ \\
\hline
0.1 & 0.25 & 0.996 & 0.992 \\
0.1 & 0.30 & 1.000 & 1.000 \\
0.1 & 0.35 & 1.004 & 1.007 \\
\hline
0.3 & 0.25 & 0.989 & 0.978 \\
0.3 & 0.30 & 1.000 & 1.000 \\
0.3 & 0.35 & 1.011 & 1.022 \\
\hline
0.5 & 0.25 & 0.982 & 0.965 \\
0.5 & 0.30 & 1.000 & 1.000 \\
0.5 & 0.35 & 1.017 & 1.034 \\
\hline\hline
\end{tabular}
\end{center}
\medskip
\label{table:compare}
\begin{minipage}{\hssize}
The fraction $f^{\Omega_\rmm=0.30}_{E(z)}$ is defined
as the ratio of the expansion function in a given cosmology to that in
$\Omega_\rmm=0.30$ cosmology, i.e., $E(z,\Omega_\rmm)/E(z,\Omega_\rmm=0.30)$.
The fraction $f^{\Omega_\rmm=0.30}_{\chi(z)}$ is defined as the ratio of the
comoving distance in $\Omega_\rmm=0.30$ cosmology, to that in another cosmology
i.e., $\chi(z,\Omega_\rmm=0.30)/\chi(z,\Omega_\rmm)$.
\end{minipage}
\end{table}

\renewcommand{\tabcolsep}{0.5cm}
\begin{table}
\caption{Cosmological dependence of comoving distance and the critical density}
\begin{center}
\begin{tabular}{ccccc}
\hline\hline
$z_\rml$ & $z_\rms$ & $\Omega_\rmm$ & $f^{\Omega_\rmm=0.3}_{\chi(z_\rml)}$ & $f^{\Omega_\rmm=0.3}_{\Sigma_{\rm crit}(z_\rml,z_\rms)}$ \\
\hline
0.1 & 0.5 & 0.25 & 0.996 & 1.0077 \\
0.1 & 0.5 & 0.30 & 1.000 & 1.0000 \\
0.1 & 0.5 & 0.35 & 1.004 & 0.9925 \\
\hline
0.1 & 0.7 & 0.25 & 0.996 & 1.0077 \\
0.1 & 0.7 & 0.30 & 1.000 & 1.0000 \\
0.1 & 0.7 & 0.35 & 1.004 & 0.9926 \\
\hline
0.1 & 0.9 & 0.25 & 0.996 & 1.0077 \\
0.1 & 0.9 & 0.30 & 1.000 & 1.0000 \\
0.1 & 0.9 & 0.35 & 1.004 & 0.9926 \\
\hline
\hline
0.5 & 0.8 & 0.25 & 0.982 & 1.0362 \\
0.5 & 0.8 & 0.30 & 1.000 & 1.0000 \\
0.5 & 0.8 & 0.35 & 1.017 & 0.9676 \\
\hline
0.5 & 1.0 & 0.25 & 0.982 & 1.0358 \\
0.5 & 1.0 & 0.30 & 1.000 & 1.0000 \\
0.5 & 1.0 & 0.35 & 1.017 & 0.9680 \\
\hline
0.5 & 2.0 & 0.25 & 0.982 & 1.0341 \\
0.5 & 2.0 & 0.30 & 1.000 & 1.0000 \\
0.5 & 2.0 & 0.35 & 1.017 & 0.9698 \\
\hline
\hline
\end{tabular}
\end{center}
\medskip
\label{table:compare_esd}
\begin{minipage}{\hssize}
The fraction $f^{\Omega_\rmm=0.30}_{\chi(z)}$ is defined the same way as in
Table~\ref{table:compare}. The fraction $f^{\Omega_\rmm=0.30}_{\Sigma_{\rm
crit}(z_\rml,z_\rms)}$ is defined as the ratio of the critical surface density
in the $\Omega_\rmm=0.30$ model to that in another cosmology, i.e.,
$f^{\Omega_\rmm=0.30}_{\Sigma_{\rm crit}(z_\rml,z_\rms)}={\Sigma_{\rm
crit}(z_\rml,z_\rms,\Omega_\rmm=0.30)}/{\Sigma_{\rm
crit}(z_\rml,z_\rms,\Omega_\rmm)}$.
\end{minipage}
\end{table}

\subsection{Galaxy-galaxy lensing measurement}
\label{sec:esd}

The primary observable for the galaxy-galaxy lensing signal is the tangential
ellipticity of background galaxies in the vicinity of foreground galaxies. The
galaxy-galaxy lensing signal is often reported as the excess surface density,
$\Delta\Sigma$ by averaging the tangential component of ellipticity around an
ensemble of galaxies
\begin{equation}
\Delta \Sigma(r_\rmp)=\tilde\Sigma_{\rm crit}(z_\rml,z_\rms) \avg{\epsilon}(r_\rmp)\,,
\end{equation}
where $r_\rmp$ denotes the projected comoving separation $r_\rmp$ between the
two galaxies at the redshift of the foreground lens, and $\Sigma_{\rm
crit}(z_\rml,z_\rms)$ is a cosmology dependent factor called the critical
surface density which is defined as
\begin{equation}
\Sigma_{\rm crit}(z_\rml,z_\rms)= \frac{c^2}{4\pi G}
\frac{D_\rmA(z_\rms)(1+z_\rml)^{-2}}{D_\rmA(z_\rml,z_\rms)D_\rmA(z_\rml)} \,.
\end{equation}
Here $D_\rmA(z_\rml)$, $D_\rmA(z_\rms)$ and $D_\rmA(z_\rml,z_\rms)$ are the
angular diameter distances to the lens, the source, and between the lens and
source, respectively, and the $(1+z_\rml)^{-2}$ factor arises from the use of
comoving units.

The cosmology dependence enters the measurement of $\Delta\Sigma$ in two ways.
The first one is similar to that discussed in the previous subsection. A given
angular scale on the sky corresponds to different comoving projected scales in
different cosmological models. Since the excess surface density is also roughly
proportional to $r_\rmp^{-1}$, the percentage change in $\Delta\Sigma$ is
similar to the percentage in the comoving distances. The second effect leads to
a change in the normalization of $\Delta\Sigma$ due to the dependence of
$\Sigma_{\rm crit}$ on the cosmological parameters and depends upon both the
source and lens redshift distribution.

In Table~\ref{table:compare_esd}, we calculate the ratios of the comoving
distance and the critical surface density for three different cosmological
models and for different combinations of source and lens redshifts, to quantify
the effect it can have on the galaxy-galaxy lensing signal. The fraction
$f^{\Omega_\rmm=0.30}_{\Sigma_{\rm crit}(z_\rml,z_\rms)}$ is defined as the
ratio of the critical surface density in a cosmological model to that in the
$\Omega_\rmm=0.30$ model. The two effects go in opposite direction making the
lensing signal less sensitive to the cosmological parameters than the
clustering signal, and even though the current statistical errors are large,
these systematic dependence of the lensing signal on the fiducial cosmology can
be also seen in real data (Miyatake et al., in preparation). 

Ideally to account for the cosmological dependence of this signal, one needs to
also consider both the source and lens redshift distributions. However if the
lens redshift range is narrow enough, one can assume the lenses to be located
at a similar effective redshift, $z_{\rm eff}$. In addition, note that
$f^{\Omega_\rmm=0.30}_{\Sigma_{\rm crit}(z_\rml,z_\rms)}$ for same lens
redshift and the same cosmology is a very weakly varying function of source
redshift (see Table~\ref{table:compare_esd}). This implies that the amplitude
correction can be safely assumed to be a constant normalization shift fairly
independent of the source redshift distribution. This correction can then be
calculated at the median redshift of the source galaxy population and applied
to the model before comparing to data. In Fig.~\ref{compare_esd}, we compare the
differences in $\Delta\Sigma$ expected due to change in the fiducial
cosmological model expected from these corrections at $z\sim0.5$ and find that
the deviations can be of the order of $4-5$ percent between $\Omega_\rmm=0.25$
and $\Omega_\rmm=0.35$ models. Although small compared to errors possible with
existing data, it is important to note, that they systematically go in the
opposite direction as the clustering signal. The errors are expected to go down
to $5$ percent or better with upcoming surveys such as the Hyper Suprime-Cam
survey.


\section{Tests on real data for the projected clustering function}
\label{sec:data}

We now use galaxies from the SDSS-III Baryon Oscillation Spectroscopic Survey
project Data Release 9 \citep[hereafter BOSS;][]{Dawson2013,Ahn2012}, and
demonstrate for the case of the projected clustering measurement how well the
effects mentioned in the Section~\ref{sec:wp} capture the relevant changes to the
measurement. We analyze the projected clustering using three different
cosmological models and show that they differ by the amount expected from the
discussion in the previous section. We choose all galaxies in the northern
region of BOSS with redshifts between $z\in[0.47,0.59]$ and $M_*>10^{10.2}
\msun$ where $M_*$ denotes the stellar mass calculated using the stellar
population synthesis models by the Portsmouth group \citep{Maraston2012}. This
yields a crude and approximate volume limited sample of galaxies (More et al.,
in preparation), however this particular aspect is not relevant to the results
presented here. The effective redshift of the sample is $z_{\rm eff}=0.53$.

\begin{figure*}
\centering
\includegraphics{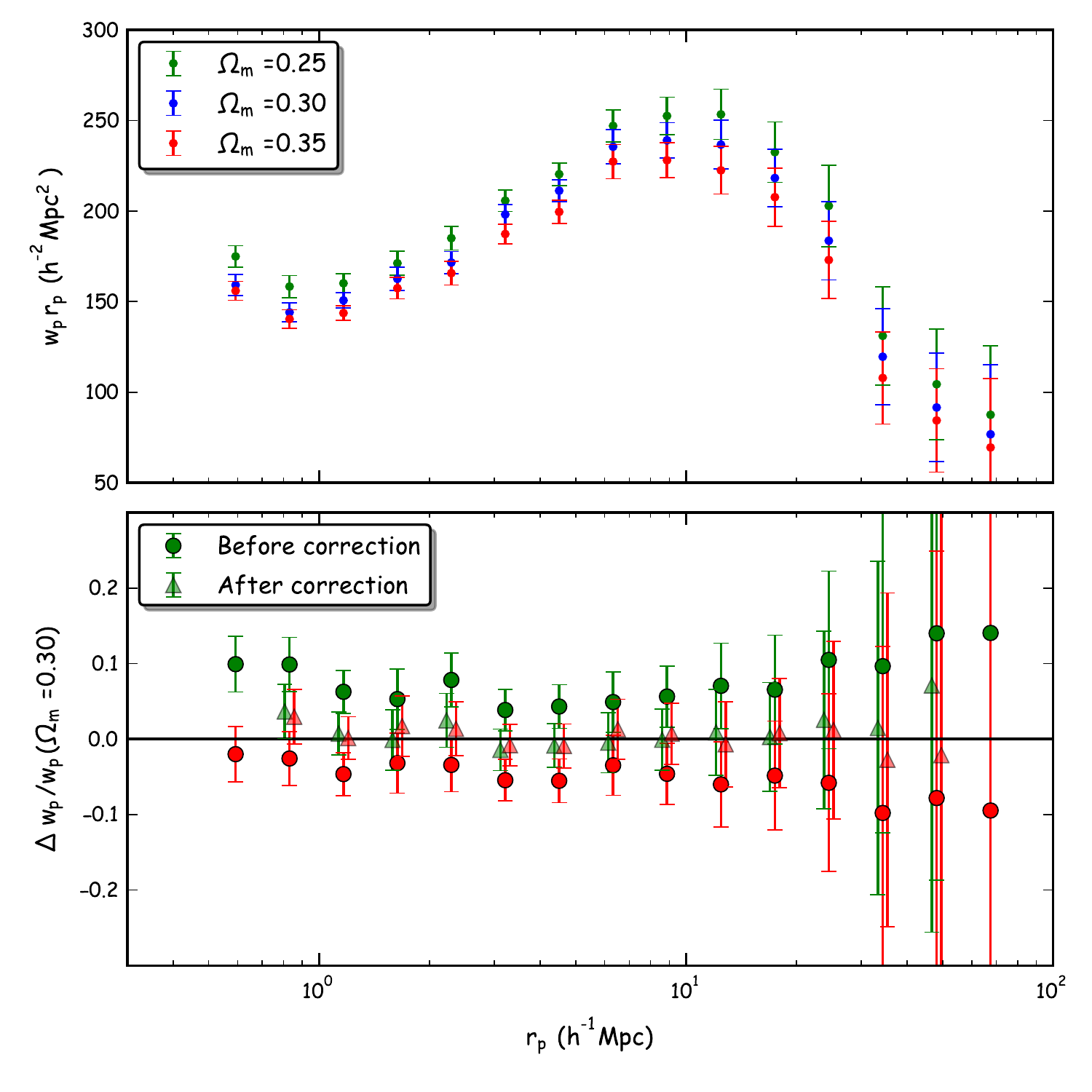}
\caption{
The projected clustering (shown as $w_\rmp\,r_\rmp$ to reduce the
dynamical scale) of an approximately volume limited subsample of CMASS
galaxies, analyzed using different flat $\Lambda$CDM cosmological models are
shown in the top panel with symbols of different colors. The clustering when
the data is analyzed with $\Omega_\rmm=0.25\,(0.35)$ is the largest (smallest).
The solid circles in the bottom panel show the difference between the
measurements with respect to the measurements in the $\Omega_\rmm=0.30$ model.
Accounting for the effects discussed in Section~\ref{sec:analytics}, we can use
the measurements performed using $\Omega_\rmm=0.25$ or $\Omega_\rmm=0.35$ model
and accurately recover the measurements in the $\Omega_\rmm=0.30$ model. The
differences between these corrected measurements and the clustering in the
$\Omega_\rmm=0.30$ model are shown using filled triangles. There is a small
offset added in the x-direction in the lower panel for clarity.
}
\label{fig:clust_comp}
\end{figure*}

In the top panel of Figure~\ref{fig:clust_comp}, we show the clustering of
galaxies measured by assuming three different flat $\Lambda$CDM cosmological
models with varying $\Omega_\rmm$ while converting the angular positions and
redshifts to distances. The solid circles in the lower panel denote the
differences between the measured clustering signal in a given cosmological
model to that in the $\Omega_\rmm=0.30$ model. The measurement errorbars are
small enough that the differences between the cosmological models are larger
than the statistical error and are systematic in nature. Fitting a constant to
the (inverse variance-weighted) residuals results in
$0.062\pm0.005\,(-0.042\pm0.003)$ for the $\Omega_\rmm=0.25\,(0.35)$ model.
\footnote{With the catalogs that use DR11, which is an internal data release in
the BOSS collaboration, the differences are even more statistically
significant.}

Next we take the measurements in the $\Omega_\rmm=0.25$ model, and predict the
clustering expected in the $\Omega_\rmm=0.30$ model as follows. To account for
the first effect discussed in Section~\ref{sec:wp}, we change the projected
comoving scale of the measurement from the $\Omega_\rmm=0.25$ analysis to
\begin{equation}
r_\rmp^{\rm corr}=r_\rmp(\Omega_\rmm=0.25)\frac{\chi(z_{\rm eff},\Omega_\rmm=0.30)}{\chi(z_{\rm eff},\Omega_\rmm=0.25)}\,.
\end{equation}
In addition we also change the amplitude of $w_\rmp$ such that
\begin{equation}
w_\rmp^{\rm corr}=w_\rmp(\Omega_\rmm=0.25)\frac{E(z_{\rm eff},\Omega_\rmm=0.25)}{E(z_{\rm eff},\Omega_\rmm=0.30)}\,.
\vspace{0.1cm}
\end{equation}
We also use similar corrections to the clustering measurements in the
$\Omega_\rmm=0.35$ model to deduce the clustering measurements in the
$\Omega_\rmm=0.30$ model. The filled triangles in the bottom panel of
Figure~\ref{fig:clust_comp} show the difference between these corrected
clustering measurement and the projected clustering measurement in the
$\Omega_\rmm=0.30$ model using filled triangles (the green [red] triangles
correspond to the $\Omega_\rmm=0.25\,[0.35]$ model corrected to that in
$\Omega_\rmm=0.30$). We see that this recovers the clustering measurement very
accurately. A constant model fit to the residuals now gives
$0.004\pm0.005\,(0.005\pm0.004)$ for the $\Omega_\rmm=0.25\,(0.35)$ model and
the residuals no longer have either just positive or negative sign. 

Although in our case, we have corrected the data for the cosmological
dependence, in modelling applications, it is better to account for the
differences in the model itself. Errors, typically done using a jack-knife
estimator (with regions of equal areal coverage), will not depend upon the
cosmological model. But errors obtained using covariances with mock simulations
populated with galaxies, may require a revision too. Exploring these details is
beyond the scope of this short letter.

\section{Summary}
\label{sec:summary}

We have presented analytical estimates for the cosmological dependence of the
galaxy luminosity function, $\Phi(M)$, the projected clustering measurement,
$w_\rmp(r_\rmp)$, and the galaxy-galaxy lensing signal $\Delta\Sigma$ obtained
from a galaxy redshift survey. We showed that these measurements change in
different cosmological models due to the difference in the comoving distances,
$\chi(z)$, the expansion functions, $E(z)$, and the change to the critical
surface density between lens and source galaxies used to measure the lensing
signal. 

These changes are small for low redshift surveys such as SDSS-I, but given the
systematic nature of the shifts can bias the cosmological constraints obtained
from a joint analysis of  $\Phi(M)$, $w_\rmp(r_\rmp)$ and $\Delta\Sigma$. These
systematic effects can be very important at higher redshifts which use these
observables and for ongoing and future surveys which can measure these
observables with ever-increasing precision. Performing a cosmological analysis
with these observables requires one to account for the cosmological parameter
dependence of the observables themselves. We have presented an analytical
framework to change the predictions for a particular cosmological model to the
ones in the fiducial cosmological model, thus allowing a fair comparison
between the model and the data.

We tested the framework for the specific case of the projected clustering
measurement $w_\rmp(r_\rmp)$ using existing data from the SDSS-BOSS survey. We
analyzed this data in the context of three different flat $\Lambda$CDM
cosmological models. We found that the differences in the measurements are
systematic in nature and significant given the errorbars. We also found that
the nature of these differences is of the same magnitude as that predicted from
the analytical framework, and hence can be easily accounted for. 

If a cosmological analysis is run without accounting for these systematic
issues, one runs the risk of biasing the cosmological parameters to values close
to the ones assumed in the fiducial cosmology used to carry out the
measurements, and significantly underestimate the errors. In future work, we
plan to quantify how the recent cosmological constraints obtained using joint
fits to abundance, clustering and lensing of galaxies by
\citet[][]{Cacciato2013} may be affected due to these systematic issues. We
also plan to investigate the effects of these systematics on the joint analysis
of clustering and lensing on large scales in \citet[][]{Mandelbaum2013},
especially for the high redshift sample.


\section*{Acknowledgments}
SM would like to acknowledge very useful and prompt inputs and comments on an
early draft version of the paper by Chris Blake, Frank van den Bosch, Marcello
Cacciato, Alexie Leauthaud and Hironao Miyatake.


\bibliographystyle{apj}
\bibliography{ref}

\end{document}